%% file: main.tex
\documentclass[sigconf]{acmart}

\input{packages}
\input{commands}

\setcopyright{acmlicensed}
\copyrightyear{2019}
\acmYear{2019}
\acmConference[ACSAC '19]{2019 Annual Computer Security Applications Conference}{December 9--13, 2019}{San Juan, PR, USA}
\acmBooktitle{2019 Annual Computer Security Applications Conference (ACSAC '19), December 9--13, 2019, San Juan, PR, USA}
\acmPrice{15.00}
\acmDOI{10.1145/3359789.3359848}
\acmISBN{978-1-4503-7628-0/19/12}

\begin{document}

\title{SIP Shaker: Software Integrity Protection Composition}
%\titlenote{Produces the permission block, and
%  copyright information}
%\subtitle{Extended Abstract}
%\subtitlenote{The full version of the author's guide is available as
%  \texttt{acmart.pdf} document}

\author{Mohsen Ahmadvand, Dennis Fischer, and Sebastian Banescu}
\orcid{1234-5678-9012}
\affiliation{%
  \institution{Technical University of Munich}
  \streetaddress{P.O. Box 1212}
  \city{Munich}
  \country{Germany}
}
\email{firstname.lastname@cs.tum.edu}

\renewcommand{\shortauthors}{Mohsen Ahmadvand et al.}

\input{abstract}



\maketitle

\input{content.tex}
\bibliographystyle{ACM-Reference-Format}
\bibliography{main}

\end{document}

\begin{tikzpicture}[auto, node distance=1.5 and 1]
	\node[present, preserved, rectangle] (address) {\texttt{address}};
	\node[present, preserved, rectangle, right=of address] (size) {\texttt{size}};
	\node[present, preserved, rectangle, rectangle, below =of address] (sc-expected) {\texttt{SC\_expected}};	
	\node[draw, rectangle, right =2 of sc-expected] (sc-computed) {\texttt{SC\_computed}};
	
	\node[rectangle, present, below =of sc-expected] (oh-hash-var) {\texttt{OH\_hash\_var}};	
	\node[rectangle, present, below =of oh-hash-var] (oh-expected-1) {\texttt{OH\_expected\_1}};	
	\node[rectangle, present, notpreserved, right =of oh-hash-var] (oh-expected-2) {\texttt{OH\_expected\_2}};	

	\draw[->,very thick] (oh-expected-2) edge node[left] {SC} (address);
	\draw[->,very thick] (oh-expected-2) edge node {SC} (size);
	\draw[->,very thick] (oh-expected-2) edge node {SC} (sc-expected);
	\draw[->,very thick] (address) edge node {SC} (sc-expected);
	\draw[->,very thick] (size) edge node {SC} (sc-expected);
	\draw[->,very thick] (oh-expected-2) edge node[left] {SC} (sc-computed);
	\draw[->,very thick] (address) edge node {SC} (sc-computed);
	\draw[->,very thick] (size) edge node {SC} (sc-computed);

	\draw[->,very thick] (address) edge [bend right=60] node[left] {OH} (oh-hash-var);
	\draw[->,very thick] (size) edge node {OH} (oh-hash-var);
	\draw[->,very thick] (sc-expected) edge node[left] {OH} (oh-hash-var);

	\draw[->,very thick] (oh-hash-var) edge node {OH} (oh-expected-1);
	\draw[->,very thick] (oh-hash-var) edge node {OH} (oh-expected-2);	

	\node[fit=(oh-expected-2),notpresent] {};
	
	\matrix [draw, right=0.5 of oh-expected-2]  {
		\node [present,label=right:Present] {}; \\
		\node [notpresent,label=right:Not Present] {}; \\
		\node [preserved,label=right:Preserved] {}; \\
		\node [notpreserved,label=right:Not Preserved] {}; \\
		\node [label=right:Dependency] {$\rightarrow$}; \\
	};
\end{tikzpicture}

\begin{tikzpicture}[auto]
	\node[rectangle, present, notpreserved] (f2) {\texttt{f2}};
	\node[present, preserved, rectangle, below =of f2] (address) {\texttt{address}};
	\node[present, preserved, rectangle, right=of address] (size) {\texttt{size}};
	\node[present, preserved, rectangle, rectangle, below =of address] (sc-expected) {\texttt{SC\_expected}};	
	\node[draw, rectangle, right =of sc-expected] (sc-computed) {\texttt{SC\_computed}};
		
	\draw[->,very thick] (f2) edge node {SC} (address);
	\draw[->,very thick] (f2) edge node {SC} (size);
	\draw[->,very thick] (f2) edge[bend right=80] node {SC} (sc-expected);
	\draw[->,very thick] (address) edge node {SC} (sc-expected);
	\draw[->,very thick] (size) edge node {SC} (sc-expected);
	\draw[->,very thick] (f2) edge node[left] {SC} (sc-computed);
	\draw[->,very thick] (address) edge node {SC} (sc-computed);
	\draw[->,very thick] (size) edge node {SC} (sc-computed);
	
	\node[rectangle, present, below =of sc-expected] (oh-hash-var) {\texttt{OH\_hash\_var}};	
	\node[rectangle, present, below =of oh-hash-var] (oh-expected-1) {\texttt{OH\_expected\_1}};	
	\node[rectangle, present, right =of oh-hash-var] (oh-expected-2) {\texttt{OH\_expected\_2}};	

	\draw[->,very thick] (address) edge [bend right=80] node {OH} (oh-hash-var);
	\draw[->,very thick] (size) edge node {OH} (oh-hash-var);
	\draw[->,very thick] (sc-expected) edge node[left] {OH} (oh-hash-var);

	\draw[->,very thick] (oh-hash-var) edge node {OH} (oh-expected-1);
	\draw[->,very thick] (oh-hash-var) edge node {OH} (oh-expected-2);	
	
	\draw[->,very thick, dashed] (oh-expected-2) edge [bend right=100] node[right] {is part of} (f2);

	\node[fit=(f2),notpresent] {};
	
	\matrix [draw, right= of sc-computed]  {
		\node [present,label=right:Present] {}; \\
		\node [notpresent,label=right:Not Present] {}; \\
		\node [preserved,label=right:Preserved] {}; \\
		\node [notpreserved,label=right:Not Preserved] {}; \\
		\node [label=right:Dependency] {$\rightarrow$}; \\
	};
\end{tikzpicture}

%% file: packages.tex
\usepackage{url}
\usepackage{xargs}                      % Use more than one optional parameter in a new commands
\usepackage[colorinlistoftodos,prependcaption,textsize=tiny]{todonotes}
\usepackage{graphicx}
\usepackage{longtable,lscape}
\usepackage{forest}
\usepackage{tikz-qtree}
\usepackage{array}
\usepackage{cleveref}
\usepackage[inline]{enumitem}   
\usepackage{booktabs}
\usepackage{multirow}
\usepackage{tabularx} 
\usepackage{collcell}
\usepackage{rotating}
\usepackage{geometry}
\usepackage{adjustbox}
\usepackage{pgfplots}
\usepackage{listings}
\usepackage{amsmath}
\usepackage{subcaption}
\usepackage{numprint}
\usepackage{makecell}
\usepackage{flushend}
\usepackage{ifpdf} 
\usepackage[nomargin,inline,marginclue,draft]{fixme}
\hypersetup{draft} 

%% file: commands.tex
\newcommandx{\unsure}[2][1=]{\todo[linecolor=red,backgroundcolor=red!25,bordercolor=red,#1]{#2}}
\newcommandx{\change}[2][1=]{\todo[linecolor=blue,backgroundcolor=blue!25,bordercolor=blue,#1]{#2}}
\newcommandx{\info}[2][1=]{\todo[linecolor=OliveGreen,backgroundcolor=OliveGreen!25,bordercolor=OliveGreen,#1]{#2}}
\newcommandx{\improvement}[2][1=]{\todo[linecolor=Plum,backgroundcolor=Plum!25,bordercolor=Plum,#1]{#2}}
\newcommandx{\thiswillnotshow}[2][1=]{\todo[disable,#1]{#2}}

\newcommand*\rot{\rotatebox{90}}
\definecolor{dkgreen}{rgb}{0,0.6,0}
\definecolor{gray}{rgb}{0.5,0.5,0.5}
\definecolor{dkgreen}{rgb}{0,0.6,0}
\definecolor{darkgreen}{RGB}{0,102,0}
\definecolor{gray}{rgb}{0.5,0.5,0.5}
\definecolor{mauve}{rgb}{0.58,0,0.82}
\crefname{lstlisting}{listing}{listings}
\Crefname{lstlisting}{Listing}{Listings}

%% file: abstract.tex
\begin{abstract}
Man-At-The-End (MATE) attackers are almighty adversaries against whom there exists no silver-bullet countermeasure.
To raise the bar, a wide range of protection measures were proposed in the literature each of which adds resilience against certain attacks on certain digital assets of a program.
Intuitively, composing a set of protections (rather than applying just one of them) can mitigate a wider range of attacks and hence offer a higher level of security.
Despite the potential benefits, very limited research has been done on the composition of protections.
Naive compositions could lead to conflicts which, in turn, limit the application of protections, raise false alarms, and worse yet, yield corrupted binaries.
More importantly, inadequate compositions of such protections are not tailored for the program at hand and thus the offered security and performance are sub-optimal.
In this paper, we first lay out a set of generic constraints for a conflict-free composition of protections. 
Then, we develop a composition framework based on a defense graph in which nodes and edges capture protections, their relations, and constraints.
The conflicts problem together with optimization requirements are then translated into a set of integer constraints. 
We then use Integer Linear Programming (ILP) to handle conflicts while optimizing for a higher security and lower overhead. 
To measure the overhead, we use a set of real-world programs (MiBench dataset and open source games). 
Our evaluation results indicate that our composition framework reduces the overhead by $\approx$ 39\% while maximizing the coverage.
Moreover, our approach yields a 5-fold decrease in overhead compared to state-of-the-art heuristics.
\end{abstract}

%% file: content.tex
\input{content/intro}
\input{content/related_work}

\input{content/requirements}

\input{content/design}

\input{content/implementation}
\input{content/evaluation}
\input{content/discussion}
\input{content/conclusions}
\input{content/availability}

%% file: content/intro.tex
\section{Introduction}\label{sec:introduction}
%% Here you describe the general motivation of this work
MATE refers to a class of adversaries whose access to the software execution environment 
allows them to monitor, subvert, and tamper with any running program in that environment~\cite{nagra2009surreptitious}. 
The consequences of such actions could range from bypassing license checks~\cite{aucsmith1996tamper}, cheating in online multi-player games~\cite{feng2008stealth}, to compromising critical infrastructures~\cite{greece2007vodafone,karnouskos2011stuxnet}.  
With the pervasiveness of software systems and the emergence of IoT in our daily life, 
the need for protecting software integrity becomes a matter of physical safety.

%% Here you describe the reason why you are not using crypto
There exists no silver-bullet countermeasure against MATE attackers. 
On the one hand, theoretically secure schemes such as indistinguishability obfuscation~\cite{garg2016candidate} are still considered to be impractical~\cite{banescu2015idea} for real-world software.
On the other hand, practical protection measures can only raise the bar against attackers
~\cite{banescu2015software,banescu2017detecting,chang2001protecting,ahmadvand2018sroh,ahmadvand2018taxonomy}.
That is, given enough time, resources, equipment, and skills, attackers eventually win~\cite{dedic2007graph}.
However, for some use cases, practical hardening measures can cause enough overhead for attackers such that defeating the protection is no longer economically attractive~\cite{ceccato2017professional}. 

%% Here you describe the narrower problem that you are trying to solve
In the past two decades, numerous protection schemes were proposed in the literature. 
Each measure essentially aims at thwarting a particular type of attack in protected programs. 
In other words, they add integrity protections on certain \emph{assets} whereby mounting certain attacks becomes relatively harder (as opposed to no applied protection). 
These assets include, but not limited to, program's code (logic), control flow, 
sensitive data (e.g. crypto-related keys), etc.~\cite{ahmadvand2018taxonomy}. 
Given the unlimited power of the MATE adversaries and the severity of compromises in any of the integrity assets, 
realizing a holistic protection with a higher coverage of different MATE attacks, is compelling.

%% Here you describe how you try to solve this problem
One way to achieve a higher degree of resilience against a wider range of attacks is to utilize a multitude of protections simultaneously.
We refer to this process as \emph{composition of protections} in the remainder of this paper.
Composing protections not only mitigates more attacks but also forms multi-layer defenses, which protect each other and are hence harder to be defeated by perpetrators.
%% Here you enumerate several challenges of the solution. I would not start a new paragraph for this
However, due to the brittle nature of integrity protection schemes, composing such protections is an open problem in literature.
On the one hand, composing oblivious hashing (OH)~\cite{chen2002oblivious} and self-checksumming (SC)~\cite{banescu2017detecting} 
could lead to a cyclic dependency between the two and ultimately could produce faulty binaries (we discuss different conflicts in detail later in \Cref{sec:requirements}).
%% I think the following sentences are redundant
%To cope with such conflicts, the state of the art proposes preemptive approaches,
%which yields not employing protections that could produce conflicts. 
%Such a naive composition, to a great extent, limits the application of protections.
On the other hand, non-cyclic dependencies are desired because code segments that are directly and/or indirectly checked (protected) by multiple integrity checks are more resilient to code patching attacks.
Furthermore, protections have different capabilities when it comes to protecting program assets. 
For instance, OH is unable to protect nondeterministic program segments, while SC does not have such a restriction~\cite{ahmadvand2018sroh}.
Moreover, the locality of the checks can have a huge impact on the imposed overhead~\cite{cappaert2008towards}. 
Simply put, carrying out expensive checks in frequently executed code segments can induce large overheads.
%% The following challenge is more or less the same as the second challenge. It seems redundant. Let's keep just one of these 2
%By the same token, the location of particular checks determines whether it can be protected by other protections or not. 
%For example, SC checks residing in nondeterministic programs segments cannot be protected by OH.
Therefore, deciding upon where and how to apply different protections in the given program can severely impact security and performance of the protected programs.
To sum it up, we believe composing a range of protection techniques can strengthen the software resilience against MATE attackers. 

\textbf{Problem.}
Naively composing protections is ineffective and 
can potentially lead to conflicts, which can severely limit the application and coverage of protections.

\textbf{Gaps.}
To the best of our knowledge (see \Cref{sec:related_work}), the gaps in literature are:
\begin{itemize}
		\item The potential conflicts arising from composing protection schemes were not thoroughly studied; 
		\item The reviewed literature makes conservative assumptions as to whether protection schemes are composable or not, 
		i.e. conflicts are prevented by forbidding the composition rather than handling them upon composition; and
		\item The existing compositions do not offer a concrete optimization methodology for better performance and security w.r.t.~the program at hand. 
\end{itemize}

\textbf{Contributions.} This paper closes the aforementioned gaps by:
\begin{itemize}
	\item Identifying three representative types of conflicts in the integrity protection composition (see \Cref{sec:requirements});
	\item Proposing a graph-based technique for protection composition, enabling the detection and handling of the conflicts at the program level  (see \Cref{sec:design});
	\item Utilizing ILP to optimize the composition for better performance and security w.r.t.~the program at hand (see \Cref{sec:design});
	\item Implementing the proposed technique as an (open source) end-to-end framework for a conflict-free composition of protection schemes  (see \Cref{sec:implementation});
	\item Empirically evaluating the effectiveness and efficiency of the proposed technique using a set of real-world programs (see \Cref{sec:evaluation}); 
\end{itemize}
\Cref{sec:threats} describes threats to validity.
\Cref{sec:conclusions} presents conclusions and ideas for future work.
%Note that the proposed framework comes with all the necessary components 
%for importing further protection schemes, adding constraints, and optimizations. 
%In \Cref{sec:implementation} we discuss these components in detail. 

%% file: content/related_work.tex
\section{Background and Related Work}\label{sec:related_work}
This paper contributes to two areas: software protection composition and protection optimization.
To introduce the area of software protection, we first review a set of representative integrity protection techniques 
that not only motivate the whole idea of the composition but are also used in our empirical evaluations.

\subsection{Integrity Protection Schemes}\label{sec:protections}

\subsubsection{Self-checksumming (SC)} SC~\cite{ghosh2013software,chang2001protecting,banescu2017detecting,ghosh2010secure,junod2015obfuscator} operates by injecting a set of guards that hash or checksum the desired (sensitive) segments of program code in memory (at runtime) and verify that the hashes match to expected values.
Because the expected values can only be known after compilation a set of placeholders are used during compilation.
An adjustment process is executed after compilation, to patch the placeholders with the actual expected values.
Consequently, if perpetrators tamper with the program code, the guards can detect them. 
Once inconsistencies are detected, a \emph{response mechanism} is triggered~\cite{nagra2009surreptitious}, 
e.g. by injecting a fault into the stack that eventually causes a crash in the application.
SC guards can further be hardened interconnecting them such that they protect each other~\cite{chang2001protecting}. 
SC guards add atypical behavior to a program, i.e. the self memory access, which exposes the guards to taint-based attacks~\cite{qiu2015identifying}.
SC is also susceptible to the memory split attacks~\cite{wurster2005generic}. 
One advantage is that SC is capable of protecting any program segment with no restriction whatsoever. 

\subsubsection{Oblivious Hashing (OH) and Short Range Oblivious Hashing (SROH)}
OH computes hashes of the program's execution trace (including memory values), by incorporating the memory reference of instructions at runtime~\cite{chen2002oblivious} and subsequently verifying them (i.e. matching them to expected values)  at random intervals. 
One way to adjust expected values is to resort to a technique similar to the SC post-patching technique~\cite{ahmadvand2018sroh}. 
%\todo[inline]{Are those post-patching things necessary or just one way of doing this?}

OH captures the effect of the program execution, not its actual code and thus cannot be deceived by attacks such as the memory split attack.
However, OH can only be applied to deterministic segments of a program, which are executed for any given input. 
This restriction severely limits the application of OH to only small parts of programs. 
SROH~\cite{ahmadvand2018sroh} partially removes this restriction by extending the coverage of OH 
to constant data residing in nondeterministic branches as well as all the branching conditions.
The input-data-dependent instructions in a program cannot be covered by SROH nonetheless. 

\subsubsection{Call Stack Integrity Verification (CSIV)}
CSIV aims to guard the access to sensitive functions in a given program~\cite{banescu2015software}. 
That is, only authentic sequences of calls can reach such functions. 
Any attempts to illegitimately jump to them triggers a response mechanism. 
To do so, a shadow stack can be created in which all (or part of) function calls on the path to sensitive functions are reflected into a hash variable. 
This can be achieved in two steps: 
\textbf{i)} accumulatively hashing random tokens~\cite{abadi2005control} in functions leading to a sensitive function; 
and \textbf{ii)} verifying the expected hashes in sensitive functions.
As opposed to SC and OH which protect the code bytes, respectively execution memory, CSIV only protects the intended control flow. 

\subsubsection{Code Mobility (CM)}
CM aims at thwarting reverse engineering by mitigating static and dynamic analyses~\cite{falcarin2011exploiting}. 
It partitions a given program into a client part and server part.
CM moves the sensitive code to the server section and places corresponding calls to the server in the client.
The server upon request delivers the sensitive code in pieces. 
This limits the amount of the code that an attacker can get access to at once.
CM imposes a constant connection between client and server. 
%The application is almost empty at the start and only contains a Binder.
%Function calls use a Binder which manages and integrates the received code blocks into the program's control-flow.
%\todo[inline]{To be honest CM as described here sounds like one would need a trusted server and a permanent connection to that server to function properly. This is a strong requirement, as opposed to the other 3 techniques presented before.}

\subsubsection{The Need for Obfuscation}
When it comes to integrity protection schemes obfuscation is no longer an additional layer of protection but a must. 
The reason being that protection guards are commonly injected into programs by instrumentation techniques.
As a direct consequence, they end up having the same syntactical resemblance.
This makes them extremely vulnerable to pattern-matching attacks. 
Plenty of proposed schemes unanimously agree on the gravity of applying obfuscation to protections~\cite{ahmadvand2018taxonomy}. 
Theoretically speaking, applying obfuscation shall not cause a conflict in protections. 
However, the application of obfuscation in practice could lead to failures, particularly  
in the adjustment processes (see \Cref{sec:obfuscation-and-scoh}). 

\subsection{Composition of Protections}
The problem of protection composition is comprised of two subproblems: 
(1) \emph{deciding on the order in which transformations shall be applied} 
and (2) \emph{deciding upon which program segments shall be protected with which protections to avoid conflicts while maximizing security and minimizing the cost}.
We discuss each of these problems in the following paragraphs.

Computing the optimal order of protections is equivalent to the problem of phase ordering in optimizing compilers and hence is undecidable~\cite{touati2006decidability}. 
Since the amenability of protection to each other can be known upfront, 
we need not to compute an optimal order of protections. 
Instead, we resort to a set of heuristics as we discuss in \Cref{sec:order-of-protections}.

The problem of applying which protections on which segments of the given program was studied in the context of software obfuscation.
Heffner et al.~\cite{heffner2004obfuscation} proposed a technique for composing obfuscation transformation by encoding dependency relations among transformations in a probabilistic \emph{Finite State Automaton (FSA)}. 
The probabilities capture different metrics as to which obfuscation is more desired to be applied. 
The composition is then achieved by a random walk on the FSA. 
Similarly, Liu et al.~\cite{liu2017stochastic} presented a technique that uses an iterative search-based algorithm to identify an optimal sequence of obfuscations over particular partitions of a given program.
Guelton et al. ~\cite{guelton2018combining} identified a set of arising conflicts amongst obfuscations provided in two open source engines (namely Tigress and O-LLVM) and clang optimization passes.
Su et al.~\cite{su2018petrinets} 
%recognize that software protection can have dependencies in which one protection could affect another one, 
%\cite{Collberg.1997} presented a sequence in which protections should be applied\cite{Collberg.1997} however, 
%lacks empirical evidence why this sequence should be used and how it affects the resulting protection of software\cite{su2018petrinets}. 
%Further, \cite{Su.2018} notice that the work from \cite{Lacey.2001}\cite{Lacey.2001} already showed that program transformations might conflict, i.e., they cannot be applied to the same program. 
%which may have consequential outcomes such as weaker protection or malfunctioning of the protected software.
proposed the design of a composition technique using Petri nets and inhibitor arcs to compute a sequence in which protections should be applied, 
in order to avoid conflicts, to a program. 
However, as we show in this paper the type of conflicts and metrics presented for obfuscation are not transferable to the software integrity protection composition problem.

ASPIRE project (\url{https://aspire-fp7.eu/}) is a framework that enables chaining of integrity protection techniques.
Nonetheless, not all of the protections can be composed due to the potential rise of conflicts~\cite{wyseur2014aspire}.
For instance, in their tool chain, code mobility and static remote-attestation shall not be utilized simultaneously.
Such preventions are hard coded in the framework within the tool configurations.
In this work, we aim to utilize potentially conflicting protections at the same time by detecting and further handling such conflicts.
Furthermore, we optimize both conflict handling, i.e. the decision on which protection shall be removed from a conflicting segment, 
and locality of protections in the program for better performance and security (coverage) of a protected program.
To achieve these goals, we model our composition problem as a graph problem and subsequently use ILP to optimally solve it. 
Our transition from expressing nodes and their relations in a graph to an ILP problem 
is inspired by the techniques introduced by Boulle et al. \cite{boulle2004compact} and Wrighton et al. \cite{wrighton2006sat}. 
%However, there are differences to this work.
%First, it targets software protections such as obfuscation and watermarking. 
%They apply their approach to a total of seven protections.

%\begin{enumerate}
%	\item Local integers merging
%	\item Watermark embedding to initialization
%	\item Opaque branch insertion
%	\item Branch inversion
%	\item Collberg-Thomborson watermarking
%	\item Function name overload
%	\item Dead code embedding
%\end{enumerate}

%While their approach allows composing these seven and other protections, the protections used in this thesis differ. 
%For example, the relationship between OH and SC could be described with a \emph{prohibition} dependency to prevent cyclic conflicts. 
%However, this composition results in applying either OH or SC. 
%Applying both protections to the same program is not possible with a \emph{prohibition} dependency. 
%Using a \emph{requirement} dependency could result in conflicts, and it cannot be used.
%
%Secondly, the proposed solution targets the dependencies between protections. 
%The programs code, control-flow, and other properties are ignored. 
%In contrast, this thesis can handle the composition of protection like OH and SC. 
%Additionally, the program to protect is taken into consideration when applying the protections.
%As a result, the thesis approach allows composing protections while optimizing for metrics such as performance, coverage, and resilience.

\subsection{Protection Optimization}
\subsubsection{Performance (Overhead)}
The locality of the guards that carry out the checks plays an important role in the imposed overhead. 
That is, if guards are placed in very frequently executed code (also called ``hot code''), e.g.~code inside nested loops, the overhead is potentially high. 
Cappert et al.~\cite{cappaert2008towards} proposed a metric to measure the \emph{hotness} of a function based on its call frequency captured in a dynamic profiling analysis.
In their protection, they favor colder functions as checkers in an attempt to induce less overhead.
In this work, we analyze code hotness at a lower level, namely basic blocks, along with other factors to optimize the locality of checkers of different types. 
%It is important to note that checkers of different protections impose different overheads.
%As an additional step in this work, we take into account the type of checkers in our decisions.
\subsubsection{Security (Coverage)}
An important metric in measuring the security of protection schemes is their 
\emph{connectivity}~\cite{ghosh2010secure,dedic2007graph,ghosh2013software} 
that captures the number of unique guards that protect a particular program element (instruction, basic blocks, function, etc). 
Understandably, achieving higher connectivity entails having more guards which (in-)directly check other guards, yielding larger overheads. 
In this work, we aim at reaching a desired connectivity level, while minimizing the overhead.
Since composed protections can potentially protect each others' guards, 
we propose a new metric as \emph{implicit coverage} that captures the program elements guards of which are protected by other guards, i.e.~indirect checking.

%% file: content/requirements.tex
\section{Requirements}\label{sec:requirements}

%\subsection{Representation for Reasoning}\label{sec:requirements:representation-for-reasoning}
%To enable automated composition of protections, we first need an abstract representation of: protection guards, 
%protected segments, and the overlaps between protections. 
%Casting protection guards and their corresponding protectees into a unified representation is the first step towards the detection of conflicts. 
%It can support reasoning about the resilience of the composed protections, for instance, 
%by measuring the number of cross-checking guards in the protected program.
%Additionally, the representation needs to be flexible enough to accommodate further knowledge about the program at hand. 
%The hotness of basic blocks and the execution flow of the program are appealing as we show later how they can be used in optimizations. 
%%Program control flow graph can particularly capture the order in which checks are hit at runtime.
%Profiling information about the program basic blocks can support the decision regarding the locality of protection guards.
%\todo[inline]{This paragraph ends abruptly. At this point I would expect to also see the representation that is used. Maybe this part could be moved to Section 4.}

\subsection{Conflict Detection}\label{sec:requirements:conflict-detection}
%Coming up with a generic approach for addressing the problem of protection composition without the knowledge of the actual conflicts is rather unattainable.
To understand the arising conflicts when combining protections, we attempted to apply the protection schemes described in \Cref{sec:protections}, in various orders on a representative dataset of programs.
In this section, we briefly report on the conflicts that we encountered in our naive composition attempts. 
From the identified conflicts we then derive a set of requirements for our composition framework.

\subsubsection{OH and SC} Both OH and SC use the concept of hashing and verification. 
OH hashes memory references, while SC hashes the process memory corresponding to a range of instructions, basic blocks or functions, at runtime. 
One can apply each of these schemes once or multiple times in arbitrary orders.
Both schemes require a finalization step (post-patching) in which the expected hash values are adjusted in the protected binary.
When OH and SC protected segments cyclically overlap, the adjustment of OH hashes invalidates previously computed SC hashes and vice versa. 
Consequently, binaries containing such conflicts would trigger false tamper-detection alarms.
This conflict in essence can occur between any pair of overlapping schemes that rely on a post-patching step.

One solution for this conflict is to resort to cross-protection placeholders (pivot bytes) like those presented in \cite{banescu2017detecting} to adjust values without invalidating others. 
However, the pivot bytes themselves can not be protected by any of the composed protections and thus they remain at risk of tampering attacks.
Alternatively, we could keep cycles intact by resorting to commutative (reversible) hash functions, which are intuitively easier to be defeated by attackers.
Therefore, we rather want to detect and subsequently prevent such cyclic conflicts in our composition. 
Simply put, \emph{deterministic finalization order} is the first requirement of our system.

\subsubsection{SC/OH and CM} Chaining CM with SC yields a different kind of conflict. 
CM mobilizes protected functions to a server from which they are retrieved only when needed and cached in a random memory address, such that the attacker does not know where to find the function.
Since the actual location of the mobilized functions shall be unpredictable, SC guards will be unable to find such functions at runtime, 
hence SC cannot protect CM protected functions.  
To avoid this type of conflict, SC needs to mark protectee segments in the code such that they cannot be removed from the code by the CM protection.
Conflicts of such can be detected by \emph{safely tracing reference changes}.
Unlike SC, OH needs not to know the address of mobilized functions. 
Applying CM on OH-protected functions as well as the inverse case impose no conflicts.

\subsubsection{Obfuscation and SC/OH}\label{sec:obfuscation-and-scoh} 
Obfuscation protects the confidentiality of data, logic and the location of code in many protection schemes.
Without obfuscation plenty of protection measures can readily be defeated by pattern matching. 
Obfuscation can conflict with protections, too. 
If it is applied after inserting the SC and/or OH guards, but before the final adjustments (when SC and/or OH are utilized), 
it may change the resemblance of placeholders introducing faults in post-patching steps.
If obfuscation is applied after final adjustments, it can break SC checks 
as hash and length of obfuscated blocks may differ from what SC guards expect.
Syntactical changes of basic blocks, their location, and lengths cause conflicts in composition of obfuscation with schemes that rely on syntactical values. 
In our system, we are in need of a \emph{safe value change tracing} mechanism to prevent such conflicts.
In other words, we need to mark certain pieces of code as unobfuscatable.

It is worthwhile to state that the identified conflicts are a complete set of possible conflicts for the given set of protections. 
That is, any other composition of the mentioned protections does not cause conflicts.
%\todo[inline]{Nothing is mentioned about CFI. Should it be removed from Section 2.1? or Should we add something?}

%% file: content/design.tex
\section{Design}\label{sec:design}
\subsection{Running Example}\label{sec:running-example}
To illustrate our design and improve readability, we use a fictional vehicle mileage counter program (\Cref{lst:running-example}) as our running example throughout this section. 
The program has two global variables: \texttt{RotationCount} and \texttt{MILEAGE}. 
The \texttt{onWheelRotationCompleted} function is an event handler that is triggered on every complete wheel rotation.
After 100 rotations, a mile is added to the car's mileage counter, which is an increase-only variable.  
The \texttt{MILEAGE} is merely altered in the \texttt{incrementMileage} function. 
In this example perpetrators aim to undermine the integrity of the mileage counter.

\lstdefinestyle{interfaces}{
  float=t
}

\begin{lstlisting}[basicstyle=\footnotesize, style=interfaces, language=C,stepnumber=1,numbers=left,numberfirstline=false,caption={Fictional car mileage counter program},label={lst:running-example}] 
int RotationCount = 0, MILEAGE = 0;
void onWheelRotationCompleted() {
  RotationCount += 1;
  if(RotationCount >= 100) {
    incrementMilage();
    RotationCount = 0;
  }
}
int incrementMileage() {
  MILEAGE += 1;
  return MILEAGE;
}
\end{lstlisting}
%void serviceCheck(int m){
%	if(m == 15000){
%		display(service 1);
%	}else if(m == 30000){
%		display(service 2);
%	}else{
%		index = m/20000;
%		if(m % 20000){
%		display(service + index);
%	}
%}
%}
\subsection{High level Process}\label{sec:high-level-process}
To achieve low coupling amongst the different protections and the composition transformation, 
%\todo[inline]{When you say system components here, do you mean the different protections or something else?}
each protection transformation stays completely unaware of the composition taking place in the framework. 
%That is, at the core of our framework are stand-alone protections. 
However, we do not want protection transformations altering the program-to-protect in one go, because conflicts might need to be settled before actually applying protections. 
To achieve this goal, we refactor transformations into \emph{two-step transformations}. 
In the first step, the transformation generates a set of proposals as to which segments it can protect in the given program
(regardless of other protections that might be part of the composition). 
Additionally, each protection submits a set of constraints (\Cref{sec:constraints}) defining under which assumptions the protection will function properly.
The second step, carries out the transformation for the requested proposals on demand. 
In order to be composable, all the protections need to be refactored to comply with this two-step transformation requirement.

Our composition framework executes the first step of protection transformations and collects their protection proposals as well as their constraints. 
The collected proposals and constraints are then merged to form \emph{protection manifests} (\Cref{sec:protection-manifest}). 
Subsequently, the created manifests are projected on a \emph{defense graph} (\Cref{sec:defense-graph}) upon which conflict analysis (\Cref{sec:conflict-handling}) and optimizations (\Cref{sec:optimizations}) are performed.

\begin{lstlisting}[basicstyle=\footnotesize,style=interfaces,language=C,stepnumber=1,numbers=left,numberfirstline=false,caption={Protected fictional car mileage counter program, note that all protections are applied at the LLVM IR level, however for the sake of simplicity our snippet is presented at the source code level},label={lst:running-example-protected}] 
int RotationCount, MILEAGE = 0;
long G = <random token>; //OH hash variable 
void onWheelRotationCompleted(){
  long h1 = <random token>;
  //overlapping SC and OH protection
  long computedSCHash=
      SC_hash(t=<AddIncrementMileage>,OH_hash(G,t), t), 
      (t=<SizeIncrementMileage>, OH_hash(G,t),t); 
  //overlapping SC and OH protection
  SC_verify(computedSCHash, (t=<expected_SC_value>, OH_hash(G,t), t));
  RotationCount += (t=1, OH_hash(G,t),t);
  if(t= RotationCount >= 100, SROH_hash(h1,t),t) {
    CSIV_Register();
    m = milageIncrement();
    RotationCount = (t=0, SROH_hash(h1,t),t);
    SROH_verify(h1, <expected_h1_value>);
  }
}
int incrementMileage(){
  CSIV_Register();
  CSIV_Verify(); //Sensitive funcions include a control flow verification
  MILEAGE += (t=1,OH_hash(G,t),t);
  OH_verify(G,<expected_oh_value>);
  G = <token> //the same token as in line 2
  return MILEAGE;
}
\end{lstlisting}

%With this, a manifest is a protection, that is, each
%protection technique like SC, OH, and SROH create multiple manifests. For example,
%SC creates one manifest each time a function is protected; OH and SROH create one
%manifest for each hash computation and hash verification. This design keeps the covered
%code changes and the number of constraints low and allows a fine-grained definition of a
%manifest.
A possible outcome of applying SC, SROH, and CSIV protections on the running example is illustrated in~\Cref{lst:running-example-protected}. In the following sections we will explain how this outcome has been obtained using our composition framework.

\begin{figure}[t!] 
	\centering
			\resizebox{0.5\textwidth}{!}{
		\includegraphics[clip, trim=1.3cm 0.5cm 0.6cm 0.5cm,width=0.5\textwidth]{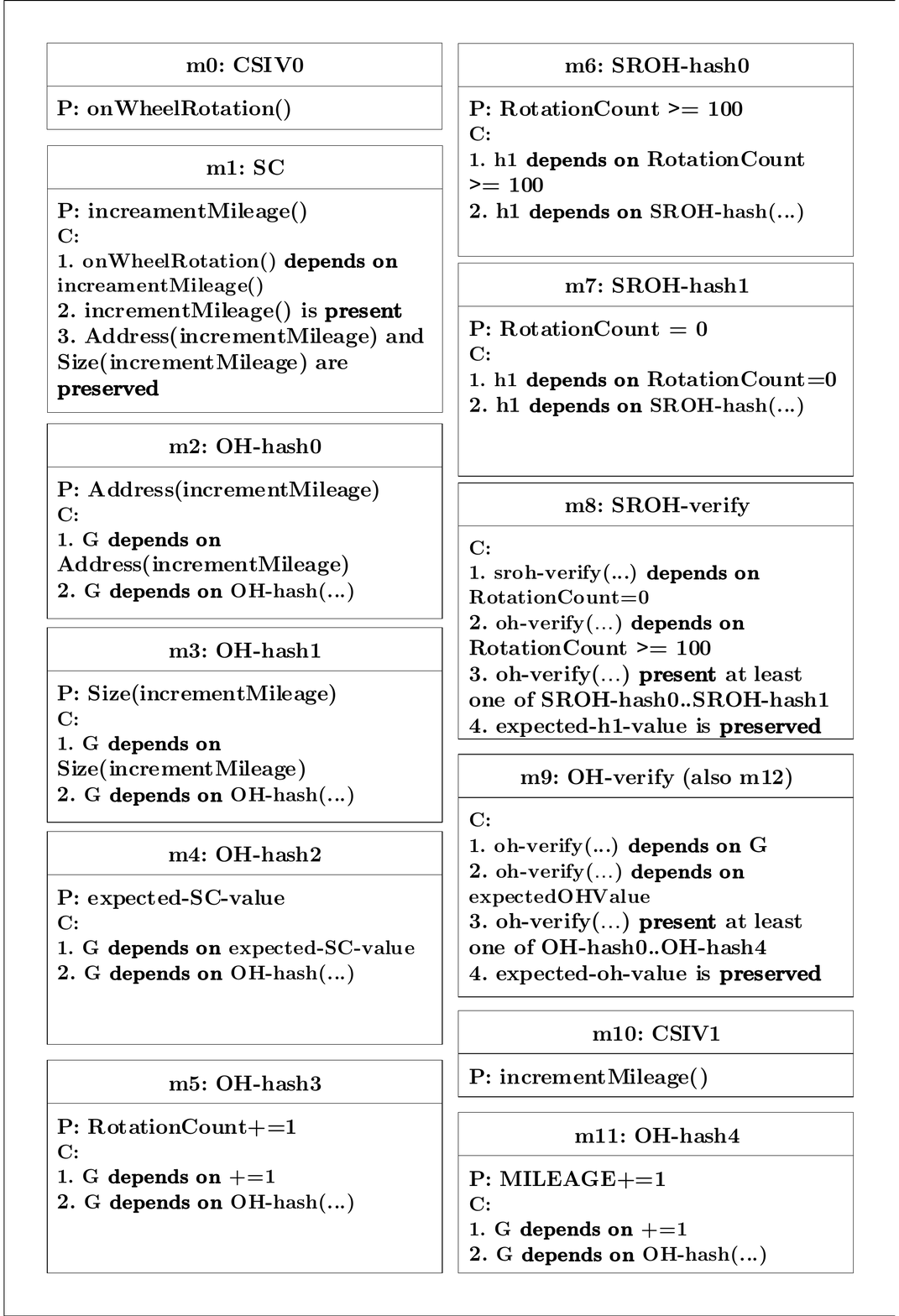}
	}
	\caption{Generated protection manifests for the example program; \texttt{P} and \texttt{C} refer to protected segments (instruction/function) and constraints, respectively}
	\label{fig:running-example-manifests}
\end{figure}

\subsection{Protection Manifest}\label{sec:protection-manifest}
A \emph{Protection Manifest} is a data structure comprising of a protection guard, a set of instructions protected by that guard, 
and constraints under which the protection behaves as intended.
Simply put, manifests reflect how and where the protection is going to be applied on the code.
Each protection technique like SC, OH/SROH, and CSIV proposes a multitude of manifests in the first step of a \emph{two-step transformation}. 

\Cref{fig:running-example-manifests} depicts 12 protection manifests output by our approach on the running example from \Cref{lst:running-example}.
Note that manifest m12 is the same as manifest m9, but they are placed in different functions.
The name of each manifest includes the protection technique acronym and optionally an appended ordinal number for the sake of unique names.
Inside the rectables representing manifests we see one or more lines prefixed by ``P''(indicating the protected instructions) and/or ``C'' (indicating the constraints, described in \Cref{sec:constraints}). 
SC and CSIV protections create stand-alone manifests, while 
OH and SROH protections represent each guard as a dependent pair of \emph{hash} and \emph{verify} manifests.
The \emph{verify} manifests do not contain any ``P'', just ``C'', because they are verifying the ``P'' from \emph{hash} manifests.
This split of \emph{verify} and \emph{hash} manifests, facilitates the cyclic dependency analysis of manifests.
%SC creates one manifest each time a function is protected; OH and SROH create one
%manifest for each hash computation and hash verification. This design keeps the covered
%code changes and the number of constraints low and allows a fine-grained definition of a
%manifest.
%\todo[inline]{This is not clear. Can you give an example of a manifest here? Otherwise the subsection is way too short.}
Each manifest keeps a reference to the protection instructions, 
which are not included in \Cref{fig:running-example-manifests} due to space limitations. 
As a direct consequence, our composition framework, 
can easily decide to drop a manifest (e.g. due to an arising conflict) resulting in the removal of the corresponding guard instructions. 

\subsection{Constraints}\label{sec:constraints}
As pointed out earlier, different protections can be applied individually to a program. 
However, each technique may yield different types of conflicts, if composed.
%In \Cref{sec:requirements} we reported on three representative conflicts that could occur in the composition of protections.
To be able to cope with those conflicts, we resort to a set of protection constraints, which are bundled up in the manifests (indicated by ``C:'' in \Cref{fig:running-example-manifests}).
Constraints are a set of rules that each protection scheme adds to their manifests. 
That is, protections will work as intended if and only if all the constraints are satisfied throughout the composition process.
%\todo[inline]{Not clear how contraints can be added by protection schemes themselves. Protection schemes are not people and cannot do anything themselves :)}
Having protection schemes specifying constraints in their manifests yields better encapsulation as the composition itself  need not to hold any scheme related knowledge.

%Every constraint is translated to an arc between the respective protection guard node and the relevant program segment nodes in the defense graph.
%\todo[inline]{Would this be just one arc or multiple?}
%We use specialized arc types for each constraint to facilitate the analysis of the defense graph. 
%A \emph{protection constraint} captures a set of conditions under which the composition produces a conflict-free result.
In our framework, we define three types of constraints, 
which directly correspond to the three requirements introduced in \Cref{sec:requirements:conflict-detection}:\\
\begin{enumerate*}
	\item  \emph{Order} constraints between manifest $m_A$, and manifest $m_B$, denote that \texttt{$m_A$} can only be finalized after \texttt{$m_B$}. 
	For example, in the m1:SC manifest from \Cref{fig:running-example-manifests}, \texttt{c1} indicates that \texttt{onWheelRotation} depends on \texttt{incrementMileage} and therefore the latter has to be finalized before.
	In a broader sense, we use the order constraints to capture dependency relationships between protection and program instructions.
	OH/SROH hash manifests use the very constraint to capture the dependency of their hash variables to hashed instructions.\\
	\item \emph{Preserve} constraints indicate that the value of certain marked instructions shall not change. 
	As depicted in \Cref{fig:running-example-manifests} preserve constraints are used by both SC and OH/SROH verify manifests to mark their placeholders as unmodifiable.\\
	\item \emph{Present} constraints between $m_A$ and $m_B$ denotes that $m_A$ can only exist if $m_B$ is present in the program. 
	SC manifests mark their protectee functions with the present constraint so they stay in the binary, e.g. \texttt{incrementMileage} is marked with such constraint in our example. 
	We use another variation of the present constraint to capture the relation between split OH/SROH hash and verify manifests. 
	Hashes that are not verified are effectively useless as they add no protection.  
	To prevent the existence of hash manifests without the corresponding verify manifests, we use present constraints such that verification manifests have to choose $n \geq 1$ out of the total number of relevant hash manifests. 
\end{enumerate*}
%\todo[inline]{IMHO this subsection should be moved to section 3, because it is not clear why these 3 types of constraints only and how you came up with them}
%Protection constraints point to a particular segment of the program-to-protect.  

\subsection{Defense Graph}\label{sec:defense-graph}
To enable automated composition of protections, we first need an abstract representation of: protection guards, 
protected segments, and the overlaps between protections. 
Casting protection guards, their corresponding protectees, and constraints into a unified representation aids detection of conflicts. 
Moreover, such a representation supports reasoning about the resilience of the composed protections, for instance, 
by measuring the number of guards that protect other guards.
Additionally, the representation needs to be flexible enough to accommodate further information about program at hand. 
Profiling information about program basic blocks can support optimization decisions regarding the location of protection guards.
%\todo[inline]{This paragraph ends abruptly. At this point I would expect to also see the representation that is used. Maybe this part could be moved to Section 4.}

We propose to use a graph-based representation as it effectively meets all our needs. 
%(\Cref{sec:requirements:representation-for-reasoning}) of our composition framework.
%\todo[inline]{Move Section 3.1 here, because I already forgot the requirements.}
The composition pass populates the graph from the proposed manifests and protection protectee relations between program sub-structures and protections.
Since protections are applied at different level of granularity (e.g. SC is applied at the function level, while OH/SROH is employed at the instruction level) graph nodes can refer to functions, instructions, or manifests. 
%Nodes represent program elements (instructions, basic blocks and functions) in constraints as well as manifests.  
%including protection guards as they are in essence program elements that are injected into the program-to-protect.
Solid black arcs represent \emph{dependency} (order) constraints.
Dashed arcs represent \emph{present} contraints over multiple elements, i.e.~constraints comprised of two or more elements are translated to arcs between their corresponding nodes in the graph.
Whilst, constraints over single elements (namely preserve and present constraints) are added to their corresponding nodes as attributes.
It is important to bear in mind that constraints are transitive. That is, if a function has incoming/outgoing arcs, all of its instructions will inherit the same relation (arcs).  
\Cref{fig:running-example-graph} depicts a simplified graph for the running example. 
The gray arcs capture the transitive arcs, i.e.~inherited dependency relations from functions to their instructions. For the sake of brevity, we omitted arcs that are deemed to be less interesting.
As illustrated in the figure the SC manifest leverages from an additional protection by three OH hash manifests (namely \texttt{OH\_hash0}, \texttt{OH\_hash1}, and \texttt{OH\_hash2}).
%Furthermore, we extend the arcs with the program control flow, 
%i.e. we connect arcs between nodes according to the program's control flow graph (CFG).
%In this representation, protection manifests encode a set of nodes and arcs in the graph.

%To encode further program-related information in our representation, 
%we specialize our graph arcs and nodes.
%To distinguish different protection relations, we introduce an arc type in our graph. 
%That is, protection arcs can have 4 different types: SC, OH/SROH, CSIV, and CM. 
%\todo[inline]{The previous sentence is not clear. Sounds like each manifest has its own type, i.e. there are as many types as manifests. How can a manifest introduce an arc over protected segments?}
We extend our graph nodes with a set of properties to accommodate for all the necessary meta data. 
For example, \texttt{<ExecFreq(N)>} where \texttt{N} refers to the execution frequency of the block in which the manifest resides.
We collect such data by profiling binaries and subsequently normalizing them. 
\ifpdf 
\begin{figure}[t!] 
	\centering
	\resizebox{0.4\textwidth}{!}{
		\includegraphics[clip, trim=1.2cm 0.5cm .63cm 0.5cm,width=0.5\textwidth]{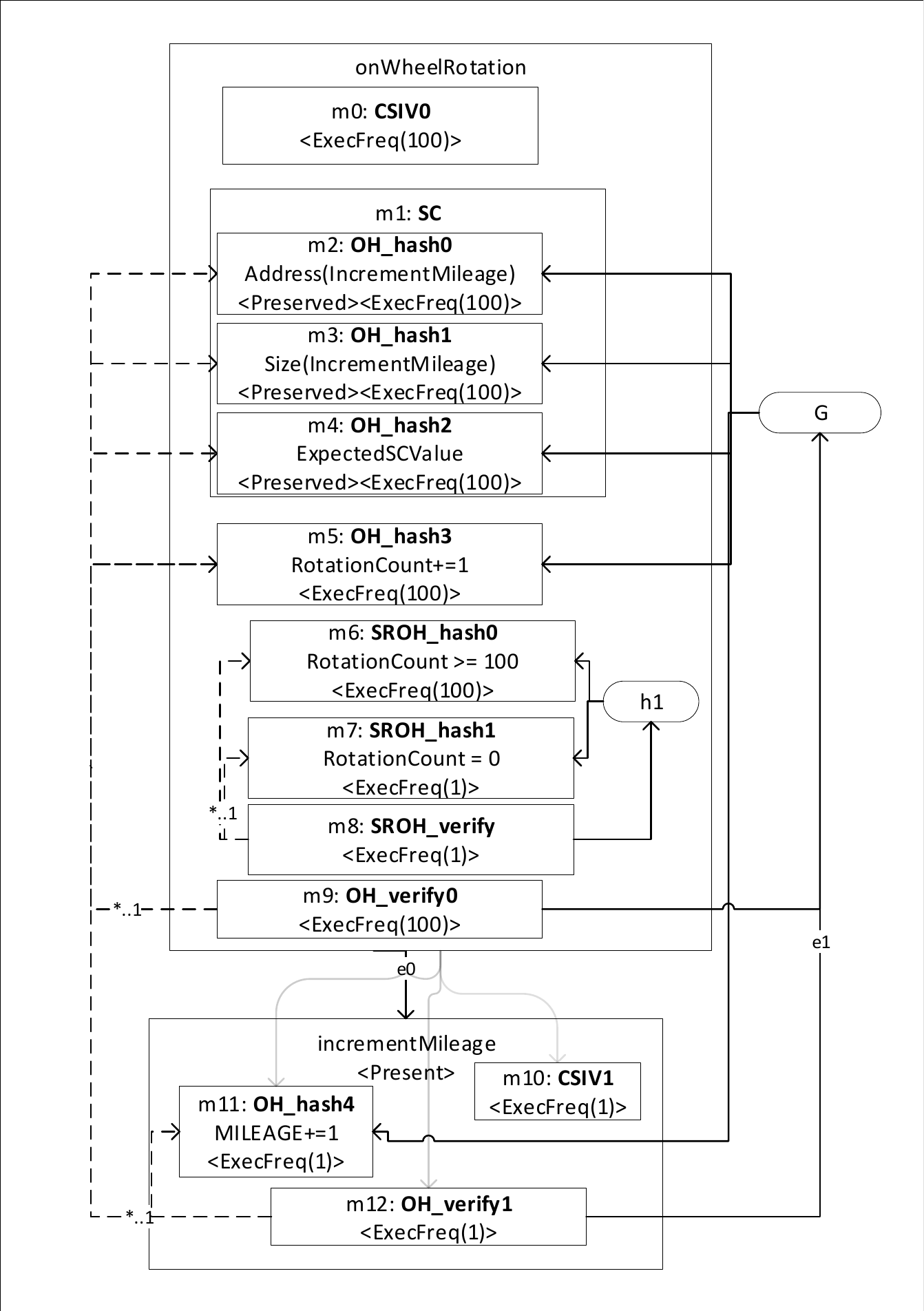}
	}
	\caption{Defense graph for the manifests of the example program; node attributes are depicted in the form of \texttt{<Attribute>}; Manifests are shown as sharp-edged rectangles while program instructions (\texttt{G} and \texttt{h1}) are shown as round-edged rectangles.}
	\label{fig:running-example-graph}
\end{figure}
\fi
%\todo[inline]{I did not understand the last sentence. Can you give an example of a small program and how a defense graph would look like for that program? Also how how a manifest look like for that program? VERY IMPORTANT: A running example to be used everywhere in section 4 would be ideal. Use different colors for different arc types.}

\subsection{Conflicts and Conflict Handling}
\label{sec:conflict-handling}
%Our constraints can be written in the form of propositional logic. 
Depending on the way that manifests are composed, \emph{order} and/or \emph{present} constraints may hold \emph{true} or \emph{false} values. 
When these types of constrains are \emph{false}, the corresponding manifests are not applied for protecting the program.
However, \emph{preserve} constraints must conserve segments from modifications, which is important to be honored by the obfuscation passes.  
Therefore, no selection of manifests violates \emph{preserve} constraints.
Manifests carrying \emph{present} constraints can only be selected if and only if their required manifests are selected as well.

%Therefore, there exist no interdependencies among constraints of such. 
%The present and preserve constraints shall be honored throughout the entire composition process including the post-patching steps.

Honoring \emph{order} constraints 
requires a delicate detection technique as a set of order constraints can collectively result in ambiguities.   
For example, assume the defense graph: $m_A$ $\rightarrow$ $m_B$ $\rightarrow$ $m_C$ $\rightarrow$ $m_A$, 
where $\rightarrow$ denotes the order constraint indicating that the left-hand manifest must be applied before the right-hand manifest. 
In the this example, a cyclic dependency materializes that renders the finalization order of the manifests impossible. 
That is, finalizing in any order results in invalidation of already finalized nodes. % (further details in \Cref{sec:conflict-resolving}).
In the defense graph from \Cref{fig:running-example-graph}, the transitive edge \texttt{e0} together with \texttt{e1} form a cycle. 
Our strategy is to break cycles by removing one or more manifests in the cycle. 
However, our framework needs to detect cycles in the first place. 
To efficiently detect cycles in the defense graph,
we utilize the strongly connected component (SCC) analysis proposed by Nuutila \cite{nuutila1994finding}. 
%Once cycles are detected, our framework decides on which manifests should be dropped from the cycles, based on the conflict resolving strategy (\Cref{sec:conflict-resolving}).
%It is important that we detect and subsequently address such ambiguities in the defense graph. 
%Conflict resolving for \emph{order} constraints consists of removing protection manifests that form cycles.
After cycle detection, there is more than one solution to the manifest removal problem. 
One out of all the manifests contributing to a cycle may randomly be removed, which would break the cycle. 
Since manifests in the graph correspond to protection mechanisms, the decision as to which manifest shall be removed entails different security and performance trade-offs.
To balance those trade-offs, we are in need of a set of security as well as performance metrics. 
Therefore, we resort to a set of metrics to measure the impact of each manifest on the security and performance overhead.
To find an optimal solution, we employ ILP to handle conflicts as described in \Cref{sec:optimizations}.

\subsection{Optimization}\label{sec:optimizations}
%There are two classes of optimizations that we can undertake: \emph{security} and \emph{performance}. 
%Here with security and performance we restrict ourself with the metrics that we introduced in \Cref{table:metrics}.

%Our technique ensures a conflict-free composition of the protections. 
%In a sense the outcome of the process is the maximum protection that composed schemes can offer without any conflicts.
%In some cases this might impose unacceptable overheads stemming from more than enough protections.  
%To balance security and performance, we propose an optimization using integer linear programming.

Our goal is to minimize the overhead of protections while attaining a set of security requirements (i.e.~explicit and implicit coverage of protections).
Running protection passes yields a set of candidate manifests, 
$M_{candidates}:\{m_0,m_1,...,m_n\}$, 
where $n$ corresponds to the total number of protection manifests, 
and $m_i \in \{0, 1\}$ for $0 \leq i \leq n$.
Precisely put, our goal is to select a subset of manifests, i.e.~ a concrete assignment of each $m_i \in M_{candidates}$, which satisfies our security requirements with the minimum possible overhead.

\subsubsection{Security Requirements}\label{sec:security-requirement-constraints}
For every manifest, the composition pass calculates the values for the following metrics:
(1) \emph{explicit instruction coverage}, which measures the instructions explicitly protected by manifests and
(2) \emph{explicit block coverage}, which measures the basic blocks explicitly protected by manifests.
Thereupon, desired security constraints can be written in the form of inequalities: $\sum_{i=1}^{n} P_{j_{m_i}} \ge D_j$ or $\sum_{i=1}^{n} P_{j_{m_i}} \le D_j$, 
where $0\le j \le p$; $p$, $P_j$, and $D_j$ correspond to the total number of scores computed for manifests, 
the $j^{th}$ score of a manifest, and the desired value for the score of interest, respectively. 
For example, \texttt{m1:SC} will be assigned with explicit instruction coverage of 6 (lines of code in \texttt{incrementMileage}), 
and explicit block coverage of 1 (there is only one basic block in \texttt{incrementMileage}).
This generic inequality can be used to have constraints over any metric value of manifests, 
except for the implicit coverage metrics (both block and instruction).
%\todo[inline]{This again needs an example because it is not clear.}

%\subsubsection{Implicit Coverage Constraints}
Implicit coverage rather depends on two (or more) manifests: a protector and a protectee. 
Therefore, it is impossible to measure the implicit coverage of a manifest without prior knowledge of other selected manifests in the solution.
For instance, \texttt{m2, m3,} and \texttt{m4}  are implicitly protecting the \texttt{incrementMileage} function as they protect the guard (\texttt{m1:SC}) that is protecting that function.  
However, their implicit coverage is equivalent to the explicit coverage of \texttt{m1} if and only if \texttt{m1} is selected in the final solution, otherwise it is 0.
%\todo[inline]{Not clear. Needs an example.}
Furthermore, there could exist multiple arcs implicitly protecting the same set of instructions. 
For example, there are three OH hash manifests (\texttt{m2}, \texttt{m3} and \texttt{m4}) that are implicitly protecting \texttt{m1}.
To precisely calculate the implicit coverage, we need to detect duplicates and subsequently consider them in our computation.

To detect duplicates, we introduce two sets of binary auxiliary variables: $e_{i,j}$ and $f_0,...,f_n$, where $0 \le i, j \le n$ and $n$ corresponds to the total number of manifests.
Every $e_{i,j} \in \{0,1\}$ indicates whether there exists an arc between $m_i$ and $m_j$ in the defense graph or not.
It is important to note that an arc ($e_{i,j}$) can only exist if and only if both of its manifests ($m_i$ and $m_j$) are present in the solution.
We formulate the desired relation by introducing a constraint such as $0 \le m_i + m_j - 2 e_{i,j} \le 1$ for all arcs.
%TODO: it is a constraint not a value, if both $m_i$ and $m_j$ exists then $e_{i,j}$ can only be 1 otherwise the constraint won't be satisfied. 
%\todo[inline]{Shouldn't this inequality be $\le 2$? Otherwise it basically prevents the case where both manifests i and j exist but there is no arc between them because that would be equal to 2.}
To deal with possible duplicate arcs, we ensure that every manifest's implicit coverage is counted only once for all the duplicate arcs.
For this purpose, we first constrain auxiliary variables $f_{i}$ as a disjunction of arcs to $m_i$, i.e. $E_i=\{ e_{j, k} | i=k\}$, $0 \le |E_i| \times f_i - \sum_{c=e_{j,i}}^{E_i}{c} \le 1$. Note that $E_i$ is a list capturing all arcs ending in $m_i$ (sink). 
In our runinng example, for $m_1$ three duplicate arcs exist: $e_{2,1}$, $e_{3,1}$ and $e_{4,1}$, which are not visible in \Cref{fig:running-example-graph}, because $m_2$, $m_3$ and $m_4$ protect $m_1$'s guard code. 
This yields a constraint such as $0 \le 3 \times f_1 - e_{2,1} - e_{3,1} - e_{4,1} \le 1$.
%\todo[inline]{It is not clear how this constrains $f_{i}$ as a disjunction of arcs to $m_i$, because it simply shows a computation $|E_i| \times f_i - \sum_{c=e_{i,j}}^{E}{e_{i,j}}$, not an inequality. I'm not sure what that even means. I understand what $E_i$ is, but what is $E$ which is used in the sum?}
Subsequently, for a precise calculation we use $f_i$ as $m_i$'s coefficient for the implicit coverage, i.e. $\sum_{0}^{n} f_i \times implicit(m_i) \le K$. 
Note that the $implicit(m_i)$ function can either return the size of implicitly protected instructions or blocks by $m_i$.
At this point we can compute solutions with the desired implicit coverage by setting $K$ to the value of interest. 

\subsubsection{Conflict Constraints}
We formulate our 3 protection constraints in the form of inequalities.
To break a cycle, an arc, from the set of arcs taking part in the cycle, needs to be removed. 
We can formulate this as $\sum_{i=1}^{C} m_i \leq C-1$, where $C$ denotes the count of manifests in a cycle. 
In our example, \texttt{m1}, \texttt{G}, and \texttt{m12} form a cyclic conflict.
A corresponding avoid-cycle constraint will limit the solver's choice to either of \texttt{m1} or \texttt{m12} manifests.
Since the objective function is subject to other security as well as performance constraints, 
the manifest with the least contribution to the desired properties will be removed. 
%\todo[inline]{I have a feeling that you are assuming that each $m_i = 1$. Is that right? This is not clear from the text. It is not clear to me why ``the manifest with the least contribution to the desired properties will be removed''}

For every set of manifests that form a cycle, we add the afore mentioned conflict constraint to the set. 
Since the SCC analysis may fail to detect all cycles, we reanalyze the outcome of the optimization to identify further (sub-)cycles. 
All the newly identified cycles are repeatedly added as constraints to the system till no more cycles are detected. 

For \emph{present} constraints, we add a constraint $m_i \le m_j$, where $i$ is the index of the manifest that cannot exist without the manifest at the index $j$.
%\todo[inline]{Not clear what the value of $m_i$ and $m_j$ are and what they represent? Are they integers? Are they vectors? they are two distinct manifests} 
It turns out that \emph{preserve} constraints do not impose any limitations on the choice of manifests as they are rather restrictions on the utilization of obfuscation on preserved placeholders. 
Therefore, we exclude them from our optimization process.
%Bear in mind that other conflicts (corresponding to present and preserve constraints) operate at the level of instructions. 

 \subsection{Order of Protections}\label{sec:order-of-protections}
 Some protections can protect the guards of other protections. 
 In such cases, a layered protection is recommended.
 The instructions explicitly protected by the protectee guards are implicitly protected by the protector guards. 
 To defeat such chained guards, attackers may need to find and disable various guards of different types, which intuitively hardens the composed protection. 
 Such chains of protections can only materialize when protections are applied in particular orders.
 Therefore, the order in which protections are applied is important.

 %defining the order of protections.
 We believe OH shall be applied after SC as SC guards can benefit from the additional protections of OH~\cite{ahmadvand2018sroh}.
% \todo[inline]{Is this consistent with OH $\rightarrow$ SC $\rightarrow$ CSIV $\rightarrow$ CM?}
 Both SC and OH can protect CSIV guards, while only OH can cover CM protected functions.
 Precisely put, the order of protection is SC $\rightarrow$ OH/SROH $\rightarrow$ CSIV $\rightarrow$ CM.
% As an alternative, we let users to define their desired orders, if they differ from the default settings. 
 It is worth noting that all obfuscations should be applied after integrity protection passes, but before finalizations, to maximize their coverage over integrity protection guards.  
 
 An interesting take is that protections can be applied multiple times. 
 In such cases, guards can potentially have a higher number of overlaps. 
 Although our design supports a multiple application of schemes,
 we consider it out of the scope of this paper. 
% \todo[inline]{can we optimize the application order of protections, for instance, to have a higher implicit protection coverage?}

%% file: content/implementation.tex
\section{Implementation}\label{sec:implementation}
All the mentioned protection schemes have an LLVM-based open source implementation.
%\url{https://github.com/tum-i22/sip-toolchain} 
For compatibility and portability reasons, we also decided to develop our composition framework in LLVM. Our entire tool chain is open source (see \Cref{sec:availability}). 
%\url{https://www.github.com/i22/sip-shaker}. 

\Cref{fig:llvm-composition} illustrates the modules in our framework along with their relation to protection and LLVM modules. 
The green boxes indicate the newly developed modules in the the system. 
Our optimizations are implemented as LLVM transformation modules. 
Protection passes (CSIV, CM, SC, and OH/SROH) are first executed which results in a set of manifests. 
The existing protection passes (CSIV, CM, SC, and OH/SROH) were re-factored to to support the two-step transformation according to \Cref{sec:high-level-process}. 
It is worthwhile to mention that supporting further protections in our framework is rather straightforward. 
To do so, a protection pass needs to be refactored to conform to our framework's two-step \emph{propose} and \emph{apply} interfaces. 

The composition pass collects the generated manifests from which it populates a defense graph. 
%TODO\todo[inline]{TODO: Fig 3 says Protection Graph not Defense Graph}
We use the lemon graph library (\url{https://lemon.cs.elte.hu}), due to its performance and flexibility, to build our defense graph with specialized arcs and nodes.  
Thereafter, a set of ILP constraints from present and order manifest-constraints are generated.
Further security requirements are also added to the problem space in the form of integer linear constraints.
We use \texttt{glpk} library (\url{https://www.gnu.org/software/glpk/}) to solve the linear problem.
Once the optimal solution is found (if security constraints are satisfiable), a conflict handling pass is engaged that remove all the manifests which are not selected by the solver.
%\todo[inline]{The figure needs to be explained a bit better. It is not clear what all the different types of arrows mean. It is not clear why the composition framework step comes after the integrity protections, but before obfuscation. I think you can remove the ``LLVM IR (optimized)'' block}
%At the bottom of \Cref{sec:high-level-process} we see that the refactored protection passes spit out a set of \emph{manifests} (represented by the dotted lines), which are used to construct a defense graph.
%Then, the conflict detection pass is executed to identify conflicts amongst proposed manifests.
%Adding new constraints is basically done by introducing new arc types in the defense graph.  
%A separate conflict handling pass is afterwards engaged, which removes problematic manifests according to the desired conflict handling strategy.
The last step is to define the finalization order of the manifests. 
This is important as the correctness of the post patching step relies on this order. 
In our framework we unified the adjustment step to follow the reverse topological patching order to avoid inconsistencies in placeholders.

%\subsection{Cost Estimation}\label{sec:implementation-cost-estimation}
\ifpdf 
\begin{figure*}[t]
	\centering
	\includegraphics[width=0.85\textwidth]{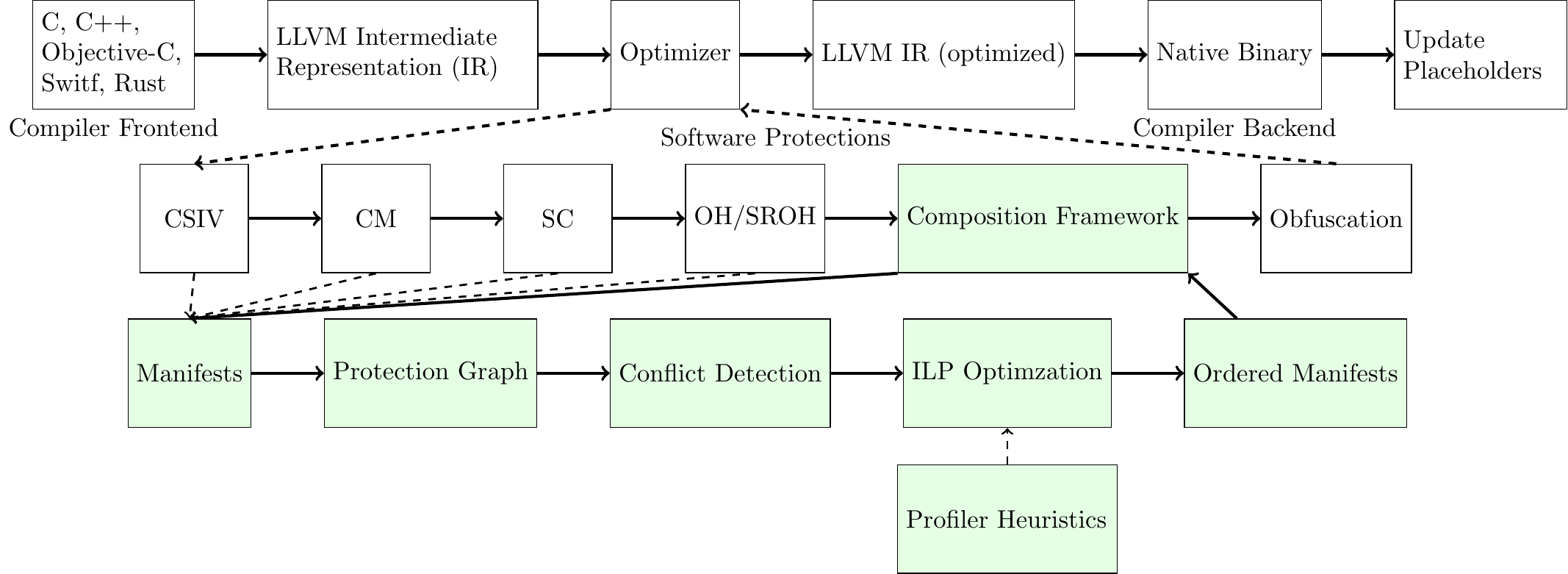}
	\caption{Module representation of our solution in LLVM framework; green boxes distinguish composition modules from LLVM and protection modules.}
	\label{fig:llvm-composition}
\end{figure*}
\fi

%\fixme{LINK TO INTRODUCTION: adding a new protection}\\
%\fixme{introducing new constraints}\\
%\fixme{adding further optimization techniques}\\

%\fixme{add further implementation details}

%% file: content/evaluation.tex
\section{Evaluation}\label{sec:evaluation}
In our evaluation, we run experiments to measure the security as well as performance of our composition. 
For performance, we benchmark the protection overhead for different coverage levels.
Moreover, we conduct an experiment in which we compare our composition optimization to a benchmark in the literature.
Regarding security, we measure protection coverage and we perform a threat analysis using attack trees.
\subsection{Dataset}\label{sec:dataset}
To evaluate effectiveness and efficiency of our solution,
we carry out a set of empirical experiments on a dataset comprised of 25 real-world programs; 
22 of these programs are taken from the MiBench dataset~\cite{guthaus2001mibench}, 
which is a representative embedded benchmark comprised of 33 programs.
%we carry out a set of empirical experiments on a dataset comprised of 29 real-world programs; 
%26 of these programs are taken from the MiBench dataset~\cite{guthaus2001mibench}, 
%which is a representative embedded benchmark comprised of 33 programs.
It is worthwhile to mention that we cannot benchmark 11 out of the 33 programs from MiBench 
10 of which is due to failures and extremely slow computations in the external pointer analysis library that SROH scheme uses~\cite{ahmadvand2018sroh}. 
The 11th program (i.e. \texttt{ispell}) appears to be faulty~\cite{ahmadvand2018sroh}.
The other three programs in our dataset are open-source games \texttt{snake} (\url{https://github.com/troglobit/snake}), \texttt{tetris} (\url{https://github.com/troglobit/tetris}) and \texttt{2048} (\url{https://github.com/cuadue/2048_game}) taken from GitHub.

Generating protected instances for six particular programs, namely \texttt{cjpeg}, \texttt{djpeg}, \texttt{say}, \texttt{susan}, \texttt{tetris} and \texttt{toast}, 
forced us to adopt a different tool.
The solver that we use in our toolchain (i.e. \texttt{glpk}) takes more than 5 minutes to find solutions for some problems. 
It becomes problematic when programs contain a large number of conflicting manifests due to subcycles. 
In such cases the composition is repeated until no cycles are detected.
For the mentioned programs the number of cycles (and subcycles) and thus the repetitions exceeds 1000 times, which makes it impractical to generate all binaries.
To cope with this problem, in the generation of protected instances of the aforementioned programs, we utilize the commercial \texttt{Gurobi} solver (\url{http://www.gurobi.com/}) under a free academic license.

%Our composition hits the timeout (i.e. 5 minutes) when protecting 7 particular programs from MiBench dataset.
%The slowdown is caused by the solver which fails to quickly find solutions for those programs. 
%We excluded those programs from our evaluations at this point. 
 
%The other three programs in our dataset are open-source games (tetris\footnote{\url{https://github.com/troglobit/tetris}}, snake\footnote{\url{https://github.com/troglobit/snake}}, and 2048\footnote{\url{https://github.com/cuadue/2048_game}} taken from github.

\subsection{Configuration of Testbed}
All performance measurements are recorded on a machine with an Intel i7-6700k processor and 32 GB of memory running the Ubuntu 18.04 LTS operating system.
For all the experiments, we use constant parameters for the SC pass (\emph{connectivity=1}) and default parameters for other protections, namely OH, SROH, CSIV and CM. 

\subsection{Coverage Optimization}
%One way to measure the effectiveness of the composed protections is to measure the coverage of different protections. 
In this evaluation we intend to measure the effectiveness of our ILP-based optimization on the coverage of the composition.
The goal is to achieve higher coverage results with lower overheads. 
For this purpose, we base our analysis on two types of coverage, namely \emph{explicit} and \emph{implicit}.
The former captures the ratio of protected instructions by the composed protections.  
The latter, however, represents the ratio of protected instructions, the guards of which are also protected by other guard(s). % guards of which instead of whose guards 
Since there exist no benchmark of the composition of our protections of interest (i.e. SC, CSIV, OH/SROH, CM), 
we build a baseline from our dataset. 
For this purpose, we use the maximum possible manifests as the baseline of our evaluation. 
In this mode, the composition is steered to select all manifests that yield no conflicts in the programs. 
We then collect the explicit and implicit coverage values achieved by the composition for every program in the the dataset. 
These values capture the maximum attainable coverages.
Subsequently, we feed those coverage values as constraints (requirements) and set the solver to find solutions with (possibly) lower overheads.
%Thereupon, we recompose the protections with the same coverage values as the baseline but this time with minimized overheads.

\Cref{table:ilp-overhead-results} presents the experiment results;
\emph{LLVM Inst. count}, \emph{Manifest all}, \emph{Explicit}, and \emph{Implicit}, \emph{Overhead\% median}, \emph{Manifest after}, and \emph{Decrease\% median} columns denote number of program instructions in LLVM IR, maximum number of applicable manifests, explicit instruction coverage, implicit instruction coverage, median of the actual overhead (after execution),
number of manifests after optimization, and percentage of the decrease over the median, respectively.
It is worthwhile to mention that the implicit/explicit instruction coverage may be larger than the total number of instructions. 
This is due to the fact that we include protection instructions (guards) in our coverage calculations. 

%However, our experiments confirm that the cost score serves as a prediction only. 

The five columns following the program name column capture the results of the maximum manifest protection (i.e. the baseline).
The other three columns starting from the \emph{Manifest after} capture the results corresponding to the ILP optimization aiming for minimizing the overhead while attaining the same explicit and implicit coverages as the baseline (maximum security).  

%The estimated cost columns shall not be confused with actual overhead numbers. Those two columns merely reflect a predicted cost value, which may or may not correlate to the actual execution overhead. We believe a precise overhead prediction depends on other environment-related factors such as caching and branch predication.
%Nevertheless, our results approves that lower estimated cost values yield less overhead in protected binaries. 
%Bear in mind that all values are rounded up to two decimal points because of which some estimated costs may appear to be equal in the table. 
%However, the optimized solutions with smaller overheads always have a lower estimated cost.

Our experiments indicate that our optimization yields an overhead decrease of 38.9\% on average.
%\Cref{table:ilp-optimization} captures the results of maximum/baseline compositions. 
%Thereupon, we conduct 3 experiments; each time setting two items to the maximum set values (feed them as constraints to ILP) while optimizing for the third item.
%That is, we attempt to find an (optimized) ILP solution (using less manifests) achieving the same
%\emph{explicit} and \emph{implicit} properties as the maximum set, but a minimized overhead.
%Similar procedure is repeated for the \emph{ilp-explicit} and \emph{ilp-implicit} rows to maximize the explicit coverage and implicit coverage while fixing the other two properties. 
%\emph{Connectivity} is another closely related metric to coverage that captures the average number of guards 
%that protect sensitive segments of a program~\cite{chang2001protecting}.
%In this experiment we aim to measure the explicit and implicit coverage along with the average connectivity of different protections. 
%\Cref{table:coverage-results-table} reports on the results of a composed protection of SC, OH/SROH, and CSIV for programs in our dataset. 
\input{content/coverage-results-table}

%\subsubsection{Coverage Optimization}
%In \Cref{sec:optimizations} we presented a technique for optimizing coverage while reducing the overhead. 
%To measure the impact of the optimization, we run an experiment in which we set the desired level of connectivity to 2. 
%The framework then optimizes the defense graph such that the desired connectivity is attained while minimizing the imposed performance penalties.
%The median improvement column in \Cref{table:coverage-results-table} reports on the optimization improvements in percentage for programs in our dataset. 

\subsection{Performance}
In the performance evaluations, we are interested in measuring 
%\textbf{a)} the overhead induced by the composition framework (defense graph construction, conflict detection and handling); and
%\textbf{b)} 
the performance improvements of the optimized composition as opposed to no-optimizations.
One way to conduct this experiment is to use randomly selected manifests as a baseline. 
Our results indicate that the presented optimization technique unsurprisingly beats such a baseline.
However, we believe random selection is unrealistic. 
In practice, compositions rather use some heuristics~\cite{ahmadvand2018sroh}.

%\subsubsection{Composition Framework}
%The efficiency of the conflict detection and handling altogether is another important concern which we address here.
%\Cref{table:conflict-detection-handling} captures the median of elapsed times in the framework on four core operations: 
%graph construction (G), conflict detection (CD), conflict handling (CH), and finally protection application (P). 
%\input{content/conflict-detection-handling-table}
%\subsubsection{Performance Optimization}
In order to capture possible improvements, we resort to a heuristic-based composition of SC and OH/SROH (presented in~\cite{ahmadvand2018sroh}) as our baseline.
Our aim is to measure the overhead difference between the optimized composition versus the heuristic one.
For this purpose, similar to the previous experiment, we feed their coverage values as constraints to our framework.
Thereafter, we use the optimization to minimize the overhead. 

To obtain comparable results, we follow their partial protection methodology in which combinations of 10\%, 25\%, 50\%, and 100\% are protected.
%To measure the impact of partial protections, we define 4 different coverage levels: 10\%, 25\%, 50\%, and 100\%. 
Each coverage level indicates the percentage of program functions that must be protected by the composition framework.
To weed out the noise from our measurements, for each coverage level we generate 20 random combinations of functions. 
%Our setup yields 2320 ($=29\times4\times 20$) protected instances of our original dataset of 29 programs.
Our setup yields 4000 protected binaries in total, 2000 ($=25\times4\times 20$) protected instances for the baseline and 2000 instances as the optimized-protected binaries.

To measure runtime overheads, we run each instance 100 times with the exact same input to the protected and unprotected instances of a program. 
For the programs in which user input is required, we pipe constant inputs by intercepting library/system calls.

\Cref{fig:performance-results} illustrates the average overhead of the composed protection for different coverage levels.
The black-margined bars represent the average overhead induced by our composition optimization technique.  
The gray-margined bars behind the black-margined bars, represent the average overhead of the heuristic-based compositions. 
Our results show that our technique on average induces 31.20\%, 41.77\%, 68.93\% and 226.98\% overhead for coverage levels of 10, 25, 50 and 100\% as opposed to 86.53\%, 166.85\%, 351.48\% and 1270.96\% (respectively) that is induced by the heuristic approach in \cite{ahmadvand2018sroh}. 
Our optimization yields a five-fold decrease in the overhead in the case of 100\% protection. 
It is noteworthy that our optimization fails to outperform the heuristic approach for some programs (or coverage levels). 
%\todo[inline]{what does ``combinations'' mean?}
Such cases indicate that there simply is no room for improvements with the requested coverage constraints. 
\ifpdf 
\begin{figure*}[!t] 
	\centering
	\includegraphics[clip, trim=0.25cm .375cm 0.3cm 0.2cm,width=0.95\textwidth]{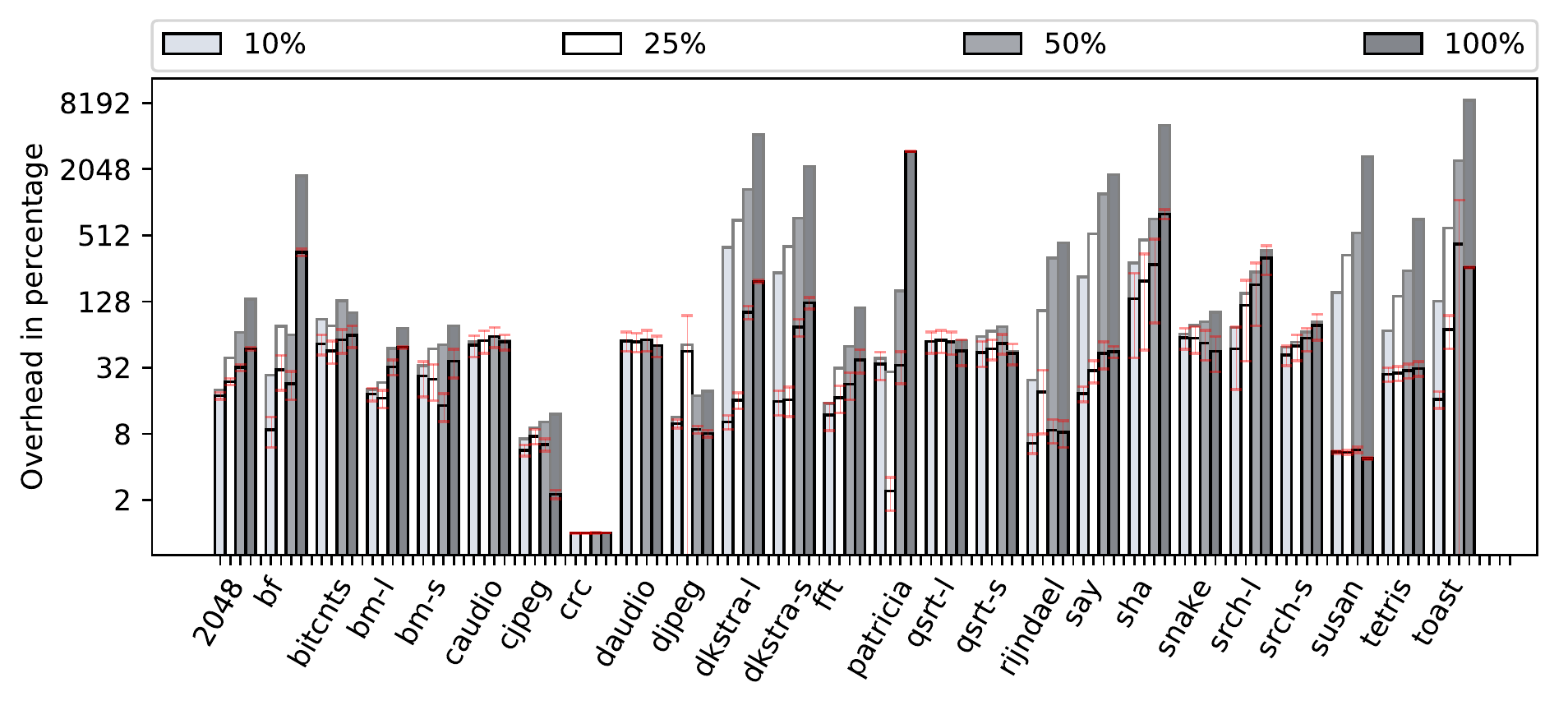}
	\caption{Performance comparison of compositions of SC and OH/SROH using heuristic-based approach vs. our optimization technique for partial protections of 10\%, 25\%, 50\%, and 100\% of program instructions; the gray-margined stacked bars capture the average overhead of the heuristic-based approach, while the black-margined bars represent the same numbers corresponding to our technique}
	\label{fig:performance-results}
\end{figure*}
\fi

%\subsection{Conflict Detection and Handling}
%In \Cref{sec:conflict-resolving} we defined constraints upon which protections work as intended. 
%Here, we are interested in ensuring that the produced binaries are conflict free. 
It is important to note that preventing conflicts was one of our main goals in this work. 
Since a majority of conflicts yield corrupted binaries, proper execution of the protected programs adds confidence to the correctness of our approach.
In our experiments all the protected binaries executed correctly (i.e. no crashes or false alarms were detected) for the given set of inputs. 
%Our experiments approve that all the protected binaries for the three conflict handling strategies execute as intended. 
%\todo[inline]{It is not clear how many protected binaries you generated from your dataset. Say this first and then explain how you checked that they executed correctly. Did you do it manually or did you write some scripts to automate this?}

\subsection{Security Analysis}\label{sec:security-analysis}
To evaluate the security of the composed protections, one has two options: 
\textbf{i)} injecting random faults (tampering) and measuring detection rates per each scheme; and 
\textbf{ii)} using attack-defense tree notation to capture the steps that an attacker has to take in order to defeat the protection. 
The assumption in the former is that perpetrators have no knowledge about the protection measure.
Subsequently, the goal is to measure what ratio of the random attacks can be detected, which is unrealistic due to 2 reasons. 
Firstly, attackers do possess or acquire knowledge about the protection techniques and thus random fault injection does not reflect the behavior of attackers in real life, as indicated in user studies~\cite{ceccato2017professional}.
Secondly, coverage metrics capture the percentage of protected instructions by each scheme.
Simply put, we already know which instructions are left unprotected or protected with weaker protections. 
Therefore, one can already know which random faults are going to go unnoticed depending on the location where they are injected.
% which renders fault injection unnecessary.
%\todo[inline]{I don't understand this second point. Can you pls rephrase?}

For the sake of simplicity, in our security evaluation we use a fictional program so-called \texttt{cool} that comes with a \emph{license check (LC)} function. 
The goal of the attacker is to use the program without any usage limitations imposed by the \texttt{LC}.
The \texttt{cool} program in various execution traces consults the \texttt{LC}. Any suspicion to violation of integrity, will be entertained with a stealthy program termination.  
Now let us assume that the \texttt{cool} program is protected using our protection composition framework. 
\Cref{fig:attack-tree} captures the steps that an attacker needs to successfully undertake in order to defeat the protected \texttt{LC}.
%\todo[inline]{ NO IDEA WHAT TO DO WITH THIS COMMENT: I wouldn't even mention the cool program. Just say any program that needs to preserve the behavior of some sensitive function, e.g.~a license check, but it can be anything else.
%The problem with the license check is that you could use symbolic execution to bypass it without tampering with the code ;)}

The root of the attack tree is \emph{defeat LC}. Our attacker has 3 means to achieve this goal: 
i) tamper with the \texttt{LC} code, e.g. to have a license which is always valid; 
ii) tamper with the callsites (usages) of \texttt{LC} function such that, for instance, they call a forged function instead of \texttt{LC};
iii) subverting the program control flow to completely bypass the \texttt{LC}.
%\todo[inline]{You have 4 nodes at depth-1, after the root. Not clear why you only give 3 means.}
All three require the attacker to first identify the \texttt{LC} function and/or its callsites in the protected program. 
The identification of the \texttt{LC} can be hardened by utilizing obfuscation techniques such as CM.
%\todo[inline]{There is no node called ``Identify the LC''}

To tamper with the logic of \texttt{LC}, the attacker (in addition to finding \texttt{LC} in the first place) may need to defeat SC and OH guards protecting \texttt{LC}. 
This, however, requires to detect SC and OH guards in the program. 
Since SC and OH guards are very amenable, it is more difficult to tackle them one by one (~\cite{ahmadvand2018sroh} discusses the difficulty of this task in depth). 
These guards are further hardened by obfuscation techniques that are applied on top of them. 

Tampering with the \texttt{LC} callsites potentially includes detection and disabling of OH/SROH/SC/CM protections in the program.
The detection is further hardened by means of applied obfuscations.

To protect the integrity of the control flow, our composition utilizes CSIV protection. 
For an attacker to \emph{defeat the CSIV protection}, it is first necessary to identify CSIV guards. 
Subsequently, the attacker needs to defeat all the potential overlapping SC/OH/SROH/CM protections.

%\todo[inline]{Some of the defese nodes in the attack tree are identical. They should be merged into 1.}
\ifpdf 
\begin{figure}[!t] 
	\centering
	\includegraphics[clip, trim=0.25cm .375cm 0.3cm 0.2cm,width=0.5\textwidth]{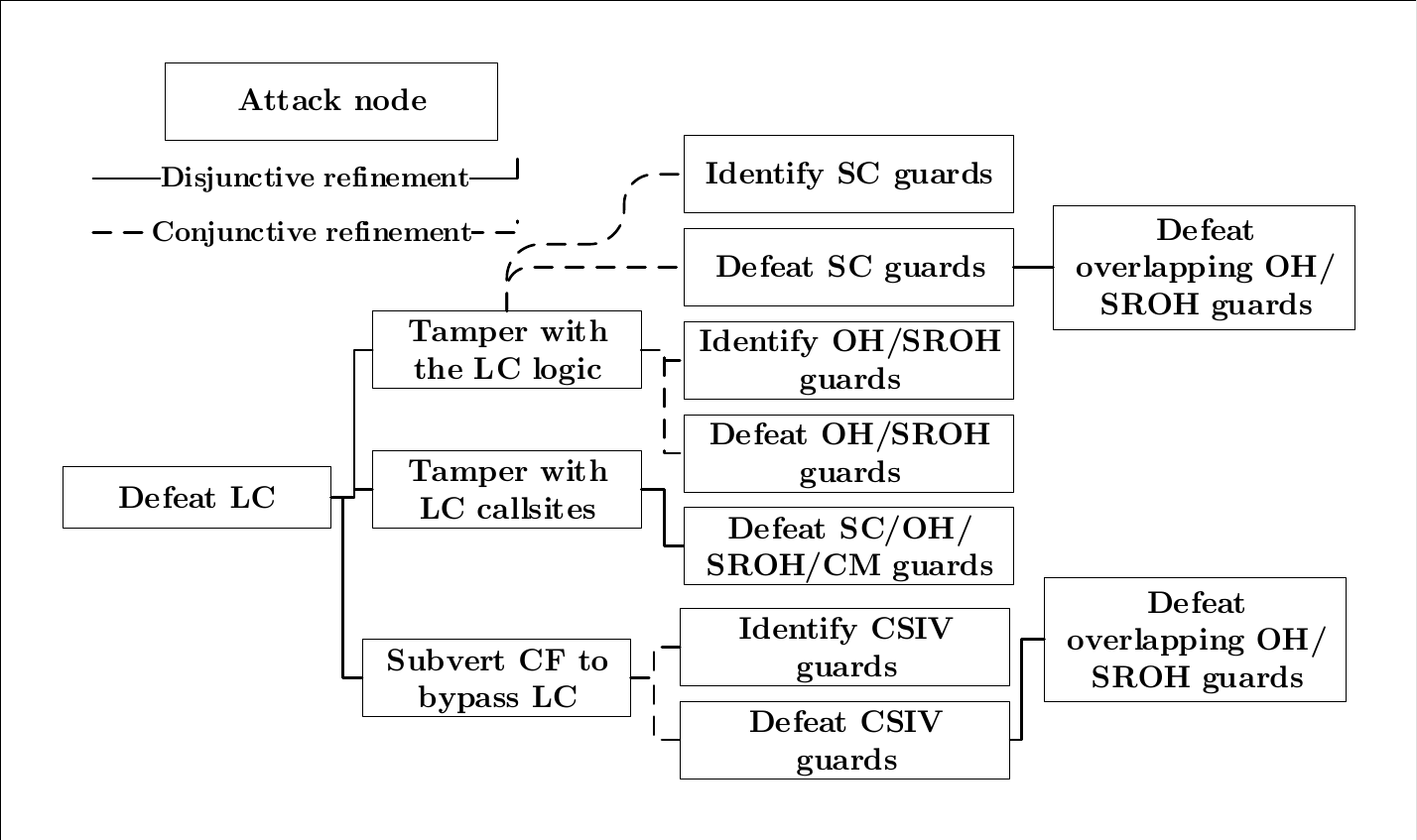}
	\caption{Security analysis of the composed protection of SC, OH/SROH, and CSIV using attack tree notation}
	\label{fig:attack-tree}
\end{figure}
\fi

%% file: content/coverage-results-table.tex
\begin{table*}[htbp]
		\resizebox{0.7\textwidth}{!}{
\begin{tabular}{|l|l|l|l|l|l|l|l|l|}
	\cline{2-9}
%	\toprule
	       \multicolumn{1}{c|}{}   &             \multicolumn{5}{c|}{Max Manifest (all possible protection)} &         \multicolumn{3}{c|}{Min Overhead}  \\\hline
%	\midrule
	\rot{\textbf{program}}              & \rot{\makecell{LLVM \\Inst. count}} & \rot{Manifest all}     & \rot{Explicit} & \rot{Implicit} & \rot{\makecell{Overhead\% \\ median}} & \rot{Manifest after}     & \rot{\makecell{Overhead\%\\ median}} & \rot{\makecell{Decrease\% \\ median}}  \\\hline
qsort\_s & 92 & 37 & 143 & 133 & 117.38 & 29 & 107.29 & 8.59 \\ \hline
crc & 147 & 105 & 569 & 561 & 12101.88 & 100 & 11454.04 & 5.35 \\ \hline
qsort\_l & 147 & 35 & 120 & 110 & 109.91 & 27 & 99.46 & 9.50 \\ \hline
djkstra\_l & 323 & 376 & 1490 & 1480 & 10155.87 & 370 & 7733.33 & 23.85 \\ \hline
djkstra\_s & 323 & 376 & 1490 & 1480 & 7424.65 & 370 & 3373.06 & 54.57 \\ \hline
caudio & 418 & 111 & 1578 & 1571 & 128.24 & 106 & 104.91 & 18.20 \\ \hline
daudio & 418 & 105 & 1556 & 1549 & 131.75 & 100 & 86.90 & 34.04 \\ \hline
bm\_s & 532 & 384 & 1284 & 1216 & 484.69 & 353 & 190.75 & 60.64 \\ \hline
tetris & 629 & 847 & 761 & 539 & 861.10 & 280 & 94.41 & 89.04 \\ \hline
bm\_l & 643 & 489 & 1395 & 1319 & 221.63 & 408 & 202.76 & 8.51 \\ \hline
sha & 657 & 1855 & 5752 & 5744 & 2581.26 & 1849 & 2401.35 & 6.97 \\ \hline
bitcnts & 664 & 157 & 1803 & 1793 & 1558.30 & 150 & 966.02 & 38.01 \\ \hline
fft & 742 & 431 & 2896 & 2885 & 2074.47 & 399 & 897.10 & 56.76 \\ \hline
2048 & 749 & 677 & 3109 & 3094 & 345.51 & 592 & 288.95 & 16.37 \\ \hline
srch\_l & 827 & 259 & 2254 & 2229 & 497.44 & 245 & 451.20 & 9.30 \\ \hline
srch\_s & 827 & 259 & 2254 & 2229 & 203.45 & 245 & 97.69 & 51.98 \\ \hline
snake & 1065 & 1826 & 6726 & 6719 & 52.17 & 1815 & 49.57 & 4.99 \\ \hline
patricia & 1087 & 431 & 2827 & 2814 & 1290.82 & 297 & 812.04 & 37.09 \\ \hline
bf & 3607 & 280 & 5489 & 5471 & 1558.30 & 210 & 473.18 & 69.63 \\ \hline
rijndael & 5866 & 418 & 7292 & 7269 & 316.90 & 345 & 88.12 & 72.19 \\ \hline
say & 6859 & 28148 & 10020 & 7902 & 30536.44 & 2698 & 4593.18 & 84.96 \\ \hline
susan & 12656 & 101599 & 6315 & 5088 & 9634.79 & 1513 & 2933.76 & 69.55 \\ \hline
toast & 13930 & 91387 & 13294 & 11501 & 17598.29 & 2215 & 3344.37 & 81.00 \\ \hline
djpeg & 52708 & 4358 & 23055 & 0 & 11427.80 & 1533 & 8534.31 & 25.32 \\ \hline
cjpeg & 54837 & 4560 & 23109 & 0 & 17394.82 & 1507 & 11101.80 & 36.18 \\       \hline
	\bottomrule
\textbf{Mean} & 6430.12 & 9580.40 & 2289.26 & 2987.84 & 5152.31 & 710.24 & 2419.18 & \textbf{38.90} \\\hline
\textbf{Median} & 742.00 & 418.00 & 1578.00 & 1793.00 & 1290.82 & 353.00 & 473.18 & \textbf{36.18} \\\hline
\textbf{Std.Dev} & 14730.39 & 26786.87 & 1859.21 & 2960.85 & 7714.51 & 784.68 & 3546.48 & 28.04 \\\hline

\end{tabular}
}
\caption{Overhead decrease of optimized solutions with maximized attainable protections (explicit and implicit coverage) with minimized overhead; columns 2-6 correspond to the maximum possible protection (applicable manifests) while columns 7-9 capture the results of optimized solutions}
\label{table:ilp-overhead-results}
\end{table*}

%% file: content/discussion.tex
\section{Threats to Validity}\label{sec:threats}

Although we have integrated the most representative integrity protections in our framework, we may have overlooked some conflicting dependency between these protection schemes.
Therefore, we are only confident with the scope of the evaluated schemes. Adding further constraints may be needed for full correctness. 
We believe the graph-based representation is sufficient for the detection of conflicts between protection schemes. 

The performance optimizations are based on the hotness of the code blocks w.r.t.~the constant inputs that were passed to the programs in the dataset. 
Further inputs, for instance, generated using symbolic execution, can be generated for the programs in an effort to visit more representative performance profile of the program after protection.

%% file: content/conclusions.tex
\section{Conclusions}\label{sec:conclusions}
Composing protections introduces layers of guards that if chosen and placed wisely can form an interconnected network of protection.
This forces attackers to defeat multiple types of protection to mount their attacks, which adds more resilience to the entire protection. 
Conflicts amongst protection and catering for the program at hand are two problems of the software protection composition.
In this work with the help of a defense graph and integer linear programming, we presented a technique that not only handles conflicts 
but also composes protections in a way that layered protections are formed. 
We further optimized the protection w.r.t. the program at hand for better performance and security.
Our evaluation results on a dataset of 25 real-world programs indicate that our composition framework decreases 
the average overhead of selected composed protections by 38.9\%
while providing the maximum security (coverage). 
Furthermore, our approach yields a five-fold decrease on the average overhead in comparison to the relevant heuristic-based approach.

As future work, we plan to empirically measure the actual resilience of composed protections as compared to single-handed protections. 
Based on the outcome of our empirical study, we aim to come up with a set of metrics to quantify the resilience of protections. 
Importing further protection schemes and identifying further constraints are also appealing.

%% file: content/availability.tex
\section{Availability}\label{sec:availability}
We made our composition framework\footnote{\url{https://github.com/mr-ma/composition-framework}} along with all protection passes, namely SC\footnote{\url{https://github.com/mr-ma/composition-self-checksumming}}, OH/SROH\footnote{\url{https://github.com/mr-ma/composition-sip-oblivious-hashing}}, and CSIV\footnote{\url{https://github.com/mr-ma/composition-sip-control-flow-integrity}}, and the interprocessor dependency analyzer \footnote{\url{https://github.com/mr-ma/composition-input-dependency-analyzer}} open source.
We also provide an artifact repository with a reproducible setting via a docker file accessible at \url{https://github.com/mr-ma/sip-shaker-artifact/}.
All our benchmarks are reproducible by means of one-step scripts provided in our evaluation repository. 
As mentioned in \Cref{sec:dataset} benchmarking 6 particular programs needs the Gurobi solver. 
We provide a comprehensive explanation as to how to generate those binaries in the artifact repository.